# Has your organization compliance with ISMS?

## A case study in an iranian Bank

## The Author

Mahsa Mohseni[1]

## Abstract

**Purpose –** The purpose of this study is proposing a model to determine the gaps between security standards requirements and the reality of implementation ISMS.
**Design/methodology/approach** – The research approach analyzes the various industry standards relevant to information security and responses gained from interviewing with 45 individuals of IT professionals and information security experts (who are chosen with targeted sampling) in order to develop a model comprising factors and subfactors which assesses compliance with ISMS (Information Security Management System) in organizations. For hypothesis test, binomial test and for ranking of factors and subfactors, Friedman test was done. This model tested in an iranian bank and the degree of compliance with ISMS calculated.

**Findings** – The case study proposes a novel model based on the standards and experience of the IT professionals and information security experts, comprising factors and subfactors which assesses the degree of readiness of an organization for implementing ISMS or the degree of compliance with this system.

**Originality/value** –Studies show Sometimes Implementing ISMS projects regarding government rules in organizations compliance with one of the existing ISMS standards are unsucceessful in achiving predifined security goals and

[1] . Master of IT management, Department of Management, Allame Tabatabaei University, Iran

Email: mahsa_mohseni@ymail.com

objectives and so waste time and other assets. This brought us in to this mind that attending to more than one standard may be inevitable.

## Introduction

The increase in hacking, the vulnerability of our computer systems and networks to physical attacks and natural disasters, the need to protect the integrity of financial accounting records and privacy and safety concerns, have resulted in the release of a number of security regulations & standards that pertain to information systems in recent years.

Information systems have penetrated every aspect of today's business processes requiring organizations to implement comprehensive solutions encompassing physical, procedural and logical forms of protection(Namjoo Choi, Dan Kim, Jahyun Goo, Andrew Whitmore, 2008)

The security issue can involve companies, medical records, military plans, phone conversations, and financial transactions.

Security describes a process of protection from any harm. Harm may indicate a loss of confidentiality, integrity, and availability. Security focuses on preventing harm resulting from both random acts of nature and intentional strategic actions (Schechter, 2004).

Armstrong (1991) admitted to the difficulty of defining the meaning of the word security in the modern business environment, and offered two meanings at the IS (information security) level: reliability (protection against accidental problems) and defensibility (protection against deliberate misuse).

Common to all information security legislation includes protecting the confidentiality, integrity and availability of information through appropriate identification and authentication procedures.

The following security controls are consistently and commonly defined in several different references (Stiller, 2010; Thompson, 2003; British Standards, 2001; Gallagher, 2003 and Bowen, 2004).

• Confidentiality - data or information is not made available or disclosed to unauthorized persons or processes.
• Integrity - data or information has not been altered or destroyed in an unauthorized manner.
• Availability - data or information is accessible and usable upon demand by an authorized person.
These three security controls are frequently referred to as the CIA Triad.

Information security means the protection of information and information systems against unauthorized access or modification of information, whether in storage, processing, or transit, and against denial of service to authorized users (free dictionary,2011). Information security includes those measures necessary to detect, document, and counter such threats. Information security is composed of computer security and communications security. Also called INFOSEC ( Webster dictionary,2011).
Information security incidents arise from many sources. They may be software and hardware engineering errors, configuration errors or inadequate physical security which allows external attackers and malicious insiders to attack the system( Finn Olav Sveen, Jose M. Sarriegi, Eliot Rich, Jose J. Gonzalez, 2007).
The provision of security in any enterprise must be tailored to that particular organization. This means that actual application of security principles depends largely on a number of factors that often vary from enterprise to enterprise (e.g., confidentiality needs for data, customers, access requirements, volatility of data value, and others) (Merrill Warkentin&Rayford B. Vaughn, 2006).
Those individuals responsible for enterprise security must be concerned with the cost of the security measures compared to it's overall strengths.
The concept of information security management system (ISMS) was first discussed during the writing and development stages of British Standard 7799 in the late 1980s. The aim of this system is implementing a kind of security controls which ensure the information security by establishing the needed infrastructure (Amir Ghotbi , Nazanin Nassir Gharechehdaghi,2012).
An information security management system (ISMS) is a set of policies concerned with information security management or IT related risks. The idioms arose primarily out of ISO 27001.The governing principle behind an ISMS is that an organization should design, implement and maintain a coherent set of policies, processes and systems to manage risks to its information assets, thus ensuring acceptable levels of information security risk (wikipedia, 2011).

## Motivation and problem statement

Sometimes Implementing ISMS projects in organizations compliance with one of the existing ISMS standards regarding government rules are unsucceessful in achieving security goals and objectives and so waste time and other assets. Therefore organizations may select two or more standards from various Industry Standards to prove compliance with the regulation that mandates its particular industry. But using two or more of this standards may waste time and money because of similarities among them.

This research investigates seven research questions:

1. How can security providers attend to all factors related to ISMS?
2. Can organizations achieve the objective of compliance with ISMS by focusing on only one Standard? If so, which standard should they address? if not which standards may be chosen and how they can be used?
3. How can security providers assess the degree of compliance with each IS factor?
4. What are the criterion for assessing the degree of compliance with ISMS?
5. What is the degree of compliance with ISMS in the bank?
6. Which are the strengthes and weaknesses of the bank in compliance with ISMS?
7. Attending to which aspects is most needed?

## Literature review

Nowadays organizations demand an acceptable level of information security. Information security management standards should certainly play a major role in this regard (Rossouw von Solms, 1999).

Standards have been developed over the years and have undergone a number of revisions.

In the security area, a number of industry standards for best practice have been released that provide guidance to companies wishing to safeguard their information and business assets.These standards cover, among other things, the requirements for information security, business continuity and risk analysis. In this paper our focus is on the third aspect of the requirements, the requirements for information security (Bonnie M. Netschert, 2008).

Four of the major industry security standards are:

(1) International Standards Organization (ISO) 17799,
(2) Control Objectives for Information and Related Technology (COBIT),
(3) the National Institute of Standards and Technologies (NIST),
(4) Information Technology Infrastructure Library (ITIL).

The ISO, COBIT, ITIL and NIST Standards all provide guidance on the major aspects of compliance with ISMS.

There is a relationship between the various laws and standards. Organizations may select from various industry standards to prove compliance with the regulation that mandates its particular industry. Healthcare organizations typically follow the guidelines defined in NIST 800-34. In contrast, financial firms and public companies typically follow COBIT to meet the Gramm-Leach-Bliley [1]and/or the Sarbanes Oxley [2]legislation. ISO-17799 is commonly used to demonstrate compliance with many other information security legislative mandates such as GISRA[3], FISMA[4]and others (Bonnie M. Netschert, 2008).

A high-level structure for each industry standard is shown in table 1 to highlight how there are many commonalities among standards.

It is apparent that all four of these Standards have common elements.

---

[1] On November 12, 1999, the ***Gramm-Leach-Bliley Act (GLBA)*** was passed by Congress. The Act requires any financial institution or business that engages in financial activities to provide a privacy notice to their customers by July 1, 2001, and when a relationship is established. GLBA applies to many types of business, including: Lending and extending credit; Providing financial or investment services; Insuring, guaranteeing, or indemnifying against loss; Underwriting or dealing with securities, Banking or closely-related banking services,etc (Aicpa, 2010).

[2]The ***Sarbanes-Oxley Act*** of 2002 (often shortened to SOX) is legislation enacted in response to the high-profile Enron and WorldCom financial scandals to protect shareholders and the general public from accounting errors and fraudulent practices in the enterprise (Bob Spurzem, 2006).

[3]***Government Information Security Reform Act***; The GISRA established information security program, evaluation, and reporting requirements for federal agencies. GISRA required agencies to perform periodic threat-based risk assessments for systems and data (Wikia, 2010).

[4]***Federal Information Security Management Act***, is a United States federal law enacted in 2002 as Title III of the E-Government Act of 2002 (Pub.L. 107-347, 116 Stat. 2899). The act recognized the importance of information security to the economic and national security interests of the United States (wikipedia, 2011).

Simplification can be obtained by understanding which requirements overlap between the COBIT; NIST; 800-34; ITIL and ISO27001. Using these results, security professionals and executives can save time and cost because they will not have to repeat the actions required for compliance with a given requirement if that requirement was already met by working with the alternant standard.

## Relevant literature

Our survey of the literature and interviews with information security experts and security providers within the security industry have not revealed the existence of any publications that provide a detailed comparison of the requirements contained in COBIT8 Version 4.0 and NIST 800- 34, ITIL and ISO 27001 standard.

The COBIT Steering Committee has completed a detailed analysis between COBIT and other Industry Standards—including the mapping between COBIT and ITIL, CMM[1], COSO[2], PMBOK[3], ISF[4] and ISO/IEC 27002 [5]to facilitate synchronization with these standards in terms of definitions and concepts.

Also an academic research (Information Security Readiness and Compliance in the Healthcare Industry) was done by Bonnie M. Netschert (2008). The research provides an in-depth analysis of security compliance regulations and
standards that pertain to the healthcare industry to achieve compliance with both the HIPAA and Sarbanes Oxley regulations.

[1] The ***Capability Maturity Model (CMM)*** is a service mark registered with the U.S. Patent and Trademark Office by Carnegie Mellon University (CMU). The Capability Maturity Model (CMM) was originally developed as a tool for objectively assessing the ability of government contractors' *processes* to perform a contracted software project (wikipedia, 2011).

[2] The **Committee of Sponsoring Organizations of the Treadway Commission** (**COSO**) is a voluntary private-sector organization, established in the United States, dedicated to providing guidance to executive management and

[3] The **Project Management Body of Knowledge** (**PMBOK)** is a collection of processes and knowledge areas generally accepted as best practice within the project management discipline. As an internationally recognised standard (IEEE Std 1490-2003) it provides the fundamentals of project management, irrespective of the type of project be it construction, software, engineering, automotive etc (Duncan Haughey, 2011).

[4] The **Information Security Forum (ISF)** is an independent, not-for-profit association of leading organizations from around the world. It is dedicated to investigating, clarifying and resolving key issues in information security, and developing best practice methodologies, processes and solutions that meet the business needs of its members (wikipedia, 2011).

[5] ***ISO/IEC 27002:2005*** comprises ISO/IEC 17799:2005 and ISO/IEC 17799:2005/Cor.1:2007. Its technical content is identical to that of ISO/IEC 17799:2005. ISO/IEC 17799:2005/Cor.1:2007 changes the reference number of the standard from 17799 to 27002. ISO/IEC 27002:2005 establishes guidelines and general principles for initiating, implementing, maintaining, and improving information security management in an organization (iso.org, 2011).

## Research model and hypotheses

After analytical research (*Phase I* of the research) explained in methodology section, a first model is gained concluding factors and subfactors related to IS. This first factors and subfactors is shown in table 2.

These are hypothesises of this research:

1. Acquisition and maintenance of information system security has a positive relation to compliance with ISMS.
   1.1. Defining Information systems Architecture has a positive relation to compliance with ISMS.
   1.2. Acquire and Maintain security Infrastructure has a posetive relation to compliance with ISMS.
   1.3. Security requirements of information systems has a posetive relation to compliance with ISMS.
   1.4. Correct processing in applications has a posetive relation to compliance with ISMS.
   1.5. Cryptographic controls has a positive relation to compliance with ISMS.
   1.6. Security of Application Software has a positive relation to compliance with ISMS.
   1.7. Security of system files has a positive relation to compliance with ISMS.
   1.8. Security in development and support Processes has a positive relation to compliance with ISMS.
   1.9. Determine Technological Direction &Technical Vulnerability Management has a positive relation to compliance with ISMS.
2. Compliance to government rules has a positive relation to compliance with ISMS.
   2.1. Compliance with legal requirements has a positive relation to compliance with ISMS.
   2.2. Compliance with security policies and standards, and technical compliance has a posetive relation to compliance with ISMS.
   2.3. technical compliance has a posetive relation to compliance with ISMS.
   2.4. Information Systems audit considerations has a posetive relation to compliance with ISMS.

3. Business Continuity Management has a posetive relation to compliance with ISMS.
    3.1. Reporting information security events and weaknesses has a posetive relation to compliance with ISMS.
    3.2. Management of information security incidents and improvements has a posetive relation to compliance with ISMS.
    3.3. Information security aspects of business continuity management has a posetive relation to compliance with ISMS.
4. Access Control has a posetive relation to compliance with ISMS.
    4.1. Business Requirement for Access Control has a posetive relation to compliance with ISMS.
    4.2. User Access Management has a posetive relation to compliance with ISMS.
    4.3. User Responsibilities has a posetive relation to compliance with ISMS.
    4.4. Network Access Control has a posetive relation to compliance with ISMS.
    4.5. Operating system access control has a posetive relation to compliance with ISMS.
    4.6. Application and Information Access Control has a posetive relation to compliance with ISMS.
    4.7. Mobile Computing and teleworking has a posetive relation to compliance with ISMS.
5. Physical and Environmental Security has a posetive relation to compliance with ISMS.
    5.1. Secure Areas has a posetive relation to compliance with ISMS.
    5.2. Equipment Security has a posetive relation to compliance with ISMS.
6. Comminucation security has a posetive relation to compliance with ISMS.
    6.1. Define and Manage Security Levels has a posetive relation to compliance with ISMS
    6.2. Operational Procedures and responsibilities has a posetive relation to compliance with ISMS.
    6.3. Third party service delivery management has a posetive relation to compliance with ISMS.
    6.4. System planning and acceptance has a posetive relation to compliance with ISMS.

6.5. Protection against malicious and mobile code has a posetive relation to compliance with ISMS.

6.6. Backup has a posetive relation to compliance with ISMS.

6.7. Network Security Management has a posetive relation to compliance with ISMS.

6.8. Media handling has a posetive relation to compliance with ISMS.

6.9. Exchange of Information has a posetive relation to compliance with ISMS.

6.10. Electronic Commerce Services has a posetive relation to compliance with ISMS.

6.11. Monitoring has a posetive relation to compliance with ISMS.

7. Organization of information security has a posetive relation to compliance with ISMS.

   7.1. Define Internal Organization & Relationships Organization has a posetive relation to compliance with ISMS.

   7.2. Define External Parties & Relationships has a posetive relation to compliance with ISMS.

8. Asset Management has a posetive relation to compliance with ISMS.

   8.1. Responsibility for assets has a posetive relation to compliance with ISMS.

   8.2. Information classification has a posetive relation to compliance with ISMS.

9. Human resources security has a posetive relation to compliance with ISMS.

   9.1. Prior to employment has a posetive relation to compliance with ISMS.

   9.2. During employment has a posetive relation to compliance with ISMS.

   9.3. Termination or change of employment has a posetive relation to compliance with ISMS.

10. Security Policy has a posetive relation to compliance with ISMS.

    10.1. Information security policy has a posetive relation to compliance with ISMS.

    10.2. Communicate Management Aims and Direction has a posetive relation to compliance with ISMS.

    10.3. Manage the IS Investment has a posetive relation to compliance with ISMS.

    10.4. Identify and Allocate Costs has a posetive relation to compliance with ISMS.

11. Education and Training has a posetive relation to compliance with ISMS.

11.1. Educate and Train Users has a posetive relation to compliance with ISMS.

11.2. Assist and Advise Customers has a posetive relation to compliance with ISMS.

## Methodology

According to (Yin 2003), the objective of case study research is to "expand and generalize theories (analytical generalization) and not to enumerate frequencies (statistical generalization)."
Because of the exploratory nature of this research, two questionnaires used to conduct quantitative surveys.
As shown in Figure 1, this outlines the three-phases of research conducted in this research.
*Phase I* of the research approach analyzes the various industry standards relevant to information security in order to determine the commonalities and differences that exist across the industry standards. These Industry Standards—ISO (International Standards Organization) 177994, NIST 800-34, COBIT Version 4.0 and ITIL are defined by specific classifications which each contain a varying number of requirements. An intensive element-by-element of the standards was conducted to uncover the differences between them.
Ultimately, the first model is designed comprising a list of factors and subfactors that could influence compliance with standards. The factors and subfactors of this first model are mentioned in the table 2.

*Phase II* (Interview Research) Questionnaire No.1 includes Likert scale, based on the first model to establish scores for each factor and to encourage critical judgment by the respondents. Delphi method was used.
This questionare was distributed among 50 security experts and IT professionals that 45 questionare was completed and returned; 22 security experts and 23 IT professionals were interviewed. So return rate gained %90. Both of these subject groups were selected based on targeted sampling.
The selected experts and professionals held from 1.5 to 20 years of experience in IT and information security. These experts were selected because of their extensive knowledge in this area.
The factors comprising this questionnaire are based on surveying the security standards that define compliance (ITIL; ISO/IEC 27002; COBIT; NIST 800-34)

There are thirty-nine factors mentioned abow in the Research model and hypotheses section.

These factors follow a Likert Scale with rating scheme in Questionnaire No.1 to determine the importance of factors and subfactors in ”compliance with ISMS” :

• 1.0 – No [viewed as not important]
• 2.0 – Minimal [ viewed as somewhat important]
• 3.0 – Average [viewed as almost important]
• 4.0 – Above average [viewed as important]
• 5.0 – Extensively [viewed as very important]

After interviewing with experts and gaining their attitudes, the structural correction was done to achive content validity. finaly from this 50 factor, 50 questions were desighned.

Finally, in *Phase III,* according to 50 finalized factors in the ultimate model, Questionnaire No.2 was designed to provide a more precise calibration of compliance with ISMS in the bank.
Survayed bank has a wide information network all over the country. The focus of this network is in the capital city and has connected the branches and departments of the bank together. The infrastructure of this network is fiber optic and also uses different technologies in connection channels.

Questionnaire No. 2 includes Likert scale with rating scheme that shows situation of the organization :

• 1.0 –No [implies no planning or action taken; or viewed as not important]
• 2.0 – Minimal [implies some planning or action taken; or viewed as somewhat important]
• 3.0 – Average [implies medium planning or action taken; or viewed as almost important]
• 4.0 – Above average [implies extensive planning or action taken; or viewed as Important]
• 5.0 –Extensively [implies very extensive planning or action taken; or viewed as very important]

The interviews with 20 IT professionals in IT department of the bank was carried out. This individuals were selected based on non-probability and convenience sampling

Last score of subfactors plus together and the last score of every factor calculates. This scores are shown on radar chart which compares present situation of organization with ideal situation.

## Data analysis

In *Phase II*, reliability analysis was done using SPSS software, α=0.92;
This number is more than 0.6 and because of it's high distance with 0.6, it is very acceptable. Now proposed model is designed.
Then factorial analysis (using principle component analysis method) was done and a new ranking of factors gained .
Conclusion of the KMO (kaiser-meryo-okin index) test was 0.704 that is more than 0.5 and shows sufficient of the sampling and fitness of data for factorial analysis.
Also the results of bartlett test that shows correlation among matrix'data that were confirmed with weight coefficient sig ۰.۰۰۰۱.
Correlation matrix shows that from 50 factors in first Questionnaire, 11 factors have eigenvalue more than1.0 and this shows validity of the questions.
Grit analysis shows eigenvalue of the factors and shows 11 factors are extracted from the questions.
Then for division of factors promax matix was used.
Afterwards hypothesis test must be done.
In this research 11 basic hypothesis and 50 Adjunct hypothesis is tested.
We used Likert Scale in Questionnaire, so non-parametric method is used for statistical analysis.
For hypothesis test, binomial test and for ranking of factors and subfactors, Friedman test was done.
Conclusions of the binomial test for basic hypothesises show that for all factors sig. Is 0.0001 (more than standard α=%5), so $H_1$ is accepted in the level of %95 and $H_0$ is rejected .
Observed proportion in hypothesises is more than 0.6 and this shows that most of the answerers are agree with impact of the factors on compliance with ISMS in organizations.
Conclusions of the binomial test for adjunct hypothesises related to each basic hypotheses show that observed proportion in adjunct hypothesises is more than 0.6 and this shows that most of the answerers are agree with impact of the subfactors on compliance with ISMS in organizations .
Conclusions of the friedman test for ranking of factors and subfactors using SPSS software is shown in tow tables. First table shows the mean of the rates of each factor/ subfactor. Second table shows statistical characteristics and $\chi^2$. In each table

**(sig.)** is compared with standard **(α=0.5)** so if **(sig.)** value is more than **(α=0.5)** so $H_0$ is accepted in the level of %95, and if **(sig.)** value is less than **(α=0.5)** then $H_0$ is rejected.
$H_0$ and $H_1$ related to rating of factors are defined below:
$H_0$: There is not important difference among factors related to compliance with ISMS.
$H_1$: There is important difference among factors related to compliance with ISMS.
$H_0$ and $H_1$ related to rating of subfactors are defined below:
$H_0$: There is not important difference among subfactors related to factor X.
$H_1$: There is important difference among subfactors related to factor X.
Final conclusions cause a new rating of factors and subfactors.
Finally we used harmonic mean For weightening each subfactor and so detemining the factors' weight.
For this objective; first, the value of each statement (that is between1 and 5 according to Likert Scale) multiply to relative abundance each of the statement gained from questionnaire No. 1; then this numbers plus together and the value of each subfactor calculates. This values normalize to gain a number between 0 and 1.0. for normalizing , the calculated value of each subfactor devides to plus of the values of all subfactors related to a factor.So the weight of each subfactor is gained.
The value of subfactors related to a factor plus together and the value of that factor calculates. For normalizing the value of each factor, this values plus together and the value of each factor devides to this number, so this values normalize. This normalized values are the weight of related factors.
Final results are shown in table 3.
In *Phase III,* reliability analysis was done using SPSS software, α=0.79; This number is more than 0.6 and acceptable.

Filled questionnaires were gained and the last score of every subfactor was calculated by multiplying the weight of each subfactor to mean score of every factor.

The Results show that the last score of every factor is 'above average' (based on an average score of 3.0) so in this areas compliance with ISMS is in a good level, unless Security Policy factor that its' score is less than 3 (=2.76) so in this area compliance with ISMS isn't in a good level .The total degree of compliance with ISMS is 3.28 that is more than 3, so the degree of compliance with ISMS in this bank is in a acceptable level.

Results of the assessing the status of the compliance with ISMS in surveyed bank is shown in table 4.

## Discussion and conclusions

In addition to the benefits provided by this study directly for the bank, a number of secondary benefits should accrue. Costumers benefit from a more secure environment as a result of security providers. The government also benefits when security providers take appropriate actions to meet legislative requirements mandated by Security Rules.

In this paper just Security Policy has average score less than 3.0 and it's compliance with ISMS is very low. Other factors have average score more than 3.0. This bank has recognized the importance of meeting standards requirements, but they have not taken the necessary actions to implement security safeguards, particularly in the area of implementing security policies.

## Summery and solutions

Non-compliance with information security carries severe civil and criminal penalties. Costly lawsuits, loss of customer confidence, embarrassment, and financial loss.

It seems that organizations aren't meeting the security demands of standards, partly because they use just one security standard in their organization.

Therefore organizations may select two or more standards from various Industry Standards to prove compliance with the regulation that mandates its particular industry. But using two or more of this standards may waste time and money because of similarities among them. That is why adherence to standards is a business decision that should be made solely by an IT organization.

So, in this paper we focus on proposing a model that contains important aspects of four of the major information security standards.

This study shows that the bank still has not compliance with some security mandates. A constraint is the lack of understanding of where to start and what actions must be taken to comply with security standards that shows weakness in Security Policy factor along with a lack of skilled IT resources, may be another critical reasons.

Business executives and IT professionals need a better understanding of strategic laws in the bank to which they may be subject in order to determine the actions that should be taken and the controls that should be implemented to set a Security Policy and build a secure environment.

## Research limitations

First, although every effort was made to ensure that the detailed comparison between the standards is correct by having an independent judge, there were areas where judgment was required and therefore some of the matches may be incorrect or misleading.

Second, the sample size used in this study was small and limited to the IT department of the surveyed bank in which case may not be generalizable to the broader population.
This research is an academic-student work and there was limitations for giving information of the bank.

## Future research

First, the efficacy of the executive framework for simplifying compliance work needs to conduct several case studies in companies attempting to apply the framework.

Second, there is a need to continuously monitor and publicize the state of compliance with ISMS in the bank industry. Such studies can inform regulators with regard to further.

Third, using fuzzy method for analysing the questionares.

Fourth, using MCDM method for ranking and weightening of factors and subfactors.

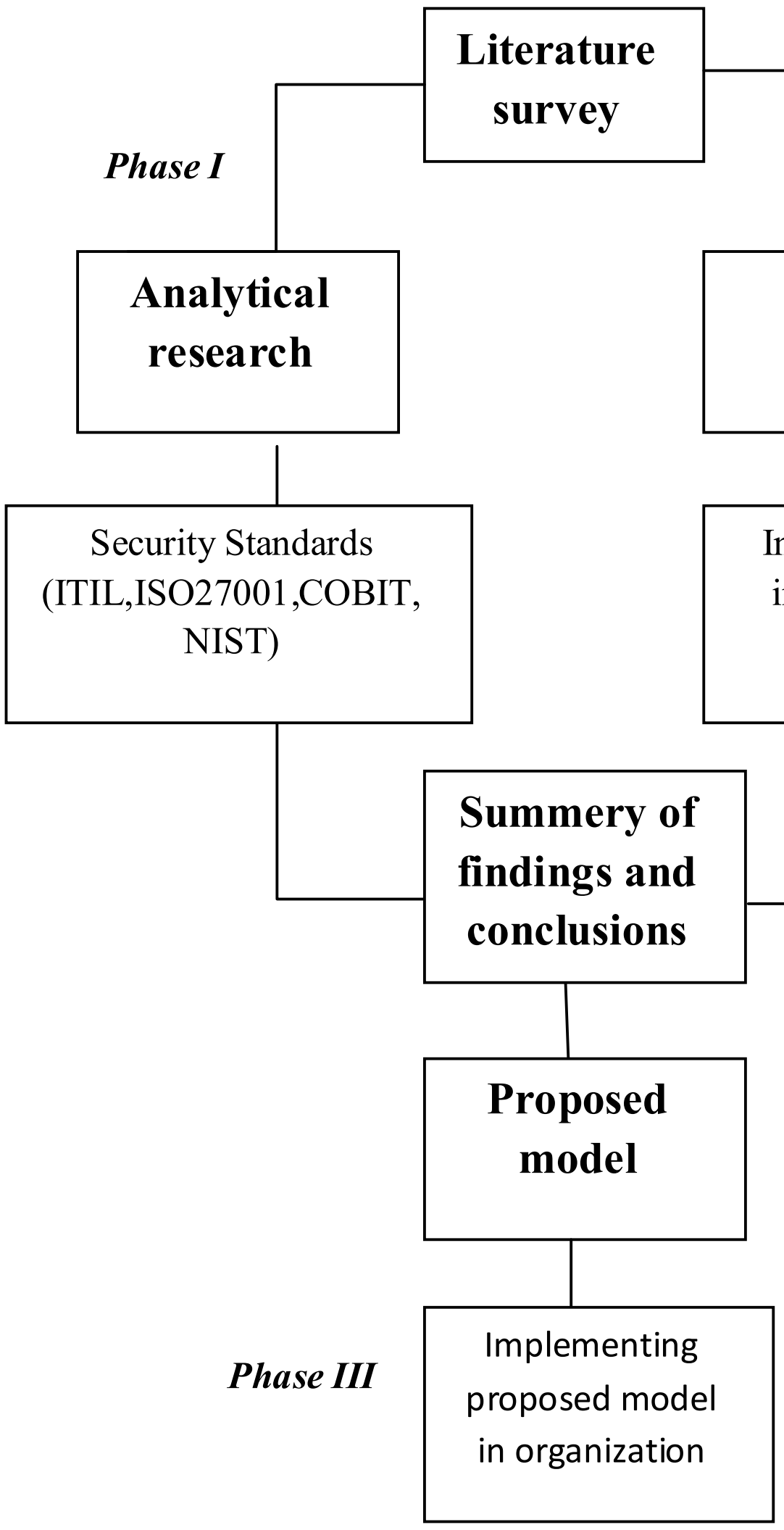

Figure (1): three-phases of research

| NIST 800-66, | COBIT 4.011 (formerly COBIT 3.0) | ITIL | ISO27001 |
|---|---|---|---|
| **Three “Functional Areas”** | **Four “Domains” (34 IT Processes)** | **Four "fundamental processes"** | **eleven" basic areas"** |
| 1. Administrative Safeguards<br>• Security Management<br>• Information Security Awareness and Training<br>• Policies and Procedures<br>2. Physical Safeguards<br>• Facility Access Controls<br>• Workstation Security<br>• Device and Media Controls<br>3. Technical Safeguards<br>• Access and Audit Controls<br>• Authentication and Transmission Security | • Planning and Organization<br>• Acquisition and implementation<br>• Deliver and Support<br>• Monitoring | • Service strategy<br>• Service design<br>• Service transition<br>• Service operation<br>• Continual service improvement | •Security Policy<br>•Organization of information security<br>•Asset Management<br>•Human resources security<br>•Physical and Environmental Security<br>•Communications and Operations Management<br>•Access Control<br>•Information systems acquisition, development and maintenance<br>•Information security incident management<br>•Business Continuity Management<br>•Compliance |

Table (1): structure for each industry standard

| **Comminucation security** | **Acquisition and maintenance of information system security** |
|---|---|
| Define and Manage Security Levels | Define the Information systems Architecture |
| Operational Procedures and responsibilities | Acquire and Maintain security Infrastructure |
| Third party service delivery management | Security requirements of information systems |
| System planning and acceptance | Correct processing in applications |
| Protection against malicious and mobile code | Cryptographic controls |
| backup | Security of system files |
| Network Security Management | Security of Application Software |
| Media handling | Security in development and support processes |
| Exchange of Information | Determine Technological Direction &Technical Vulnerability Management |
| Electronic Commerce Services | **Compliance to government rules** |
| Monitoring | A-Compliance with legal requirements |
| Manage Changes | Compliance with security policies and standards |
| **Organization of information security** | technical compliance |
| Define Internal Organization & Relationships | Information Systems audit considerations |
| Define External Parties & Relationships | **Business Continuity Management** |
| **Asset Management** | Reporting information security events and weaknesses |
| Responsibility for assets | Management of information security incidents and improvements |
| Information classification | Information security aspects of business continuity management |
| **Human resources security** | **Access Control** |
| Prior to employment | Business Requirement for Access Control |
| During employment | User Access Management |
| Termination or change of employment | User Responsibilities |
| **Education and Training** | Network Access Control |
| Educate and Train Users | Operating system access control |
| Assist and Advise Customers | Application and Information Access Control |
| **Security Policy** | Mobile Computing and teleworking |
| Information security policy | **Physical and Environmental Security** |
| Communicate Management Aims and Direction | Secure Areas |
| Manage the IS Investment | Equipment Security |
| Identify and Allocate Costs | |

Table (2): the factors and subfactors related to compliance with ISMS

| weight | Score of factor | Factors related to the compliance with ISMS |
|---|---|---|
| 0.152 | 2550.5 | Information systems security |
| 0.073 | 1232.6 | Compliance to government rules |
| 0.077 | 1290.2 | Business Continuity Management |
| 0.177 | 2970.8 | Access Control |
| 0.044 | 737.0 | Physical and Environmental Security |
| 0.244 | 4100.3 | Comminucation security |
| 0.051 | 860.7 | Organization of information security |
| 0.048 | 800.2 | Asset Management |
| 0.075 | 1250.4 | Human resources security |
| 0.025 | 422.8 | Security Policy |
| 0.033 | 552.7 | Education and Training |
| 1.000 | 16768.2 | **sum** |

Table (3): Final results

| Harmonic score | weight | Final Score of factor | Factors related to the compliance with ISMS |
|---|---|---|---|
| 0.52 | 0.152 | 3.45 | Information systems security |
| 0.24 | 0.073 | 3.29 | Compliance to government rules |
| 0.25 | 0.077 | 3.30 | Business Continuity Management |
| 0.66 | 0.177 | 3.72 | Access Control |
| 0.16 | 0.044 | 3.76 | Physical and Environmental Security |
| 0.86 | 0.244 | 3.53 | Comminucation security |
| 0.16 | 0.051 | 3.12 | Organization of information security |
| 0.16 | 0.048 | 3.30 | Asset Management |
| 0.25 | 0.075 | 3.29 | Human resources security |
| 0.07 | 0.025 | 2.76 | Security Policy |
| 0.11 | 0.033 | 3.38 | Education and Training |
| 3.28 | **The degree of compliance with ISMS** | | |

Table (4): Results of the assessing the status of the compliance with ISMS in surveyed bank